\documentclass[twocolumn,amsmath,amssymb,superscriptaddress,nofootinbib]{revtex4-1}
\pdfoutput=1

\usepackage[english]{babel}
\usepackage{amsthm}
\usepackage{amssymb}
\usepackage{amsmath}
\usepackage{amsthm}
\usepackage[]{graphicx}
\usepackage{tensor}
\usepackage{color}
\usepackage{framed}
\usepackage{framed}
\usepackage[bookmarks,linktocpage, colorlinks=true, plainpages = false, citecolor = blue,  linkcolor=blue, urlcolor = blue, filecolor = blue]{hyperref} 
\usepackage{natbib}
\usepackage{dcolumn}
\usepackage{enumerate}
\usepackage{bm}
\usepackage{graphicx}
\usepackage{caption}
\usepackage{subcaption}

\theoremstyle{definition}

\graphicspath{{./figures/}}

\usepackage{tikz}
\usetikzlibrary{decorations.pathmorphing, patterns,shapes}

\input{epsf}

\begin{document}

\allowdisplaybreaks
\begin{titlepage}

\title{
Gravitational entropy and the flatness, homogeneity and isotropy puzzles}
\author{Neil Turok}
\email{neil.turok@ed.ac.uk}
\affiliation{Higgs Centre for Theoretical Physics, James Clerk Maxwell Building, Edinburgh EH9 3FD, UK}
\affiliation{Perimeter Institute, 31 Caroline St N, Ontario, Canada}
\author{Latham Boyle}
\email{lboyle@perimeterinstitute.ca}
\affiliation{Perimeter Institute, 31 Caroline St N, Ontario, Canada}
\begin{abstract}
\vspace{.5cm}
\noindent 
We suggest a new explanation for the observed large scale flatness, homogeneity and isotropy of the universe. The basic ingredients are elementary and well-known, namely Einstein's theory of gravity and Hawking's method of computing gravitational entropy. The new twist is provided by the boundary conditions we recently proposed for  ``big bang" type singularities dominated by conformal matter, enforcing $CPT$ symmetry and analyticity. Here, we show that, besides allowing us to describe the big bang, these boundary conditions allow new gravitational instantons, enabling us to calculate the gravitational entropy of cosmologies which include radiation, dark energy and space curvature of either sign. We find the gravitational entropy of these universes, $S_g \sim S_\Lambda^{1/ 4} S_r$, where $S_\Lambda$ is the famous de Sitter entropy and $S_r$ is the total entropy in radiation. To the extent that $S_g$ exceeds $S_\Lambda$, the most probable universe is flat. By analysing the perturbations about our new instantons, we argue it is also homogeneous and isotropic on large scales. 
\end{abstract}
\maketitle
\end{titlepage}



Pictures of the earth from space reveal it to be remarkably round and smooth, with a curvature radius much larger than the distances we explore in our everyday lives. Hence, on those scales, it is often a good approximation to treat the earth's surface as flat. The explanation of this flatness involves thermodynamics. First, gravity pulls matter inwards, so its potential energy is minimised in a spherical configuration. Second, relative motions of the earth's constituent atoms generate friction: as a mountain is pulled downwards or a rock falls, gravitational potential energy is converted into heat. Even if the earth is regarded as a closed, conservative system, there are vastly more ways of distributing its internal energy among its $\sim 10^{50}$ atoms as heat than there are of creating more complicated, less spherical geometries. Taken together, gravity, friction and the earth's many atoms explain why it is locally flat~\cite{Parnell}.

In this Letter, we provide a similar, entropic explanation for the observed flatness, homogeneity and isotropy of the cosmos. Our argument rests on a new calculation of gravitational entropy, along the lines advocated by Hawking and others in the context of black hole thermodynamics~\cite{Hawking:1976de,Gibbons:1976ue,Wald}. The calculation is made possible by our new approach to the boundary conditions for cosmology, implementing  $CPT$ symmetry and analyticity at the bang, quantum mechanically, to solve many puzzles~\cite{BFT1,BFT2,BT1,BT2}. We outline the connection to Penrose's classical Weyl curvature hypothesis~\cite{Penrose} in the conclusion. Using these new boundary conditions, we showed that cosmologies with radiation, dark energy and curvature are periodic in imaginary proper time, with a Hawking temperature given by that of the corresponding de Sitter spacetime~\cite{BT1}. This is a strong hint that the solutions should be interpreted thermodynamically.  Here, in an appropriate time slicing we find the solutions describe new gravitational instantons, one for each value of the macroscopic parameters, allowing us to calculate the gravitational entropy. For a given, positive dark energy density, with radiation included the gravitational entropy can be arbitrarily large. As the total entropy is raised, the most likely universe becomes progressively flatter. Furthermore, inhomogeneous or anisotropic perturbations are suppressed. Hence, {\it to the extent that the total entropy exceeds the de Sitter value, the most probable universe is not only flat, but also homogeneous and isotropic on large scales.} 

The partition function of a statistical ensemble can be represented by a path integral over configurations that are periodic in imaginary time. Hawking and collaborators first used these methods to investigate black hole thermodynamics, a topic which has since burgeoned into holographic studies~\cite{Penington:2019npb,Almheiri:2019psf} and experimental tests using analog systems~\cite{Weinfurtner:2010nu,Kolobov:2019qfs}. We shall follow Hawking {\it et al.}'s original approach here~\cite{GHP,Hawking,SchoenYau, GibbonsHawking}.  In the cosmological setting, it is an excellent approximation to treat the radiation as a relativistic fluid, in local thermal equilibrium at a temperature declining inversely with the scale factor. The instantons we present are saddle points of the path integral for gravity: real, Euclidean-signature solutions to the Einstein equations for cosmologies with dark energy, radiation and space curvature.  The associated semiclassical exponent $iS/\hbar$ is real, large and positive, and may be interpreted as the gravitational entropy. 


One of us has given a formal argument, based on Picard-Lefschetz theory, that a path integral for a quantum gravitational {\it transition amplitude} can never yield a positive semiclassical exponent~\cite{FLT,FLT1,FLT2}.  However, for a {\it statistical ensemble}, a formal argument indicates precisely the opposite. Consider, for example, the partition function $Z(\beta) ={\rm Tr} (e^{-\beta H})$.  The time reparameterization invariance of general relativity means that the Hamiltonian $H$ vanishes on physical states~\cite{note1}.  Thus, $Z=e^{S}$ simply counts the number of states. If $Z$ is approximated by a saddle, the semiclassical exponent {\it must} be positive. 

In this Letter, we treat background spacetimes with
\begin{align}
ds^2=a^2(t)(-n^2 dt^2 +\gamma_{ij}(x)dx^i dx^j),
 \label{e0i}
\end{align}
where $n$ is the lapse, which may be set constant by reparameterizing $t$, and $a(t)$ is the scale factor. Comoving 3-space is assumed to be maximally symmetric, with metric $\gamma_{ij}(x)$ and Ricci scalar $6 \kappa$. For $\kappa>0$, it is $S^3$, with volume $V=2 \pi^2 \kappa^{-3/2}$. For $\kappa<0$, we assume a compact subspace of $H^3$, whose volume is $2 \pi^2 |\kappa|^{-3/2}$ times a topology-dependent constant. For ease of presentation, we generally leave the constant implicit. 

The partition function may be represented formally as $Z=$Tr$(\int dn e^{-i H n})$, with the integral enforcing the contraint that the total Hamiltonian $H=0$~\cite{note1}.  In the first (background) approximation, the Hamiltonian for gravity, $H_g$ depends only on the scale factor $a$ whereas, due to conformal invariance, the Hamiltonian for the radiation, $H_r$ is independent of $a$. (Using conformal time, the Hamiltonian is dimensionless.)  Hence we can trace over the radiation at some temperature $T=\beta^{-1}$, using Tr$_r (e^{-\beta H_r})= e^{-\beta F_r} = e^{S_r-\beta U_r}$ where $S_r$ is the total entropy in the radiation and $U_r$, expressed as a function of $S_r$, is the associated energy in the conformal frame. Analytically continuing to Lorentzian time, $\beta=i \,n$, with $n$ real, we obtain $Z=e^{S_r}$Tr$_g(\int dn e^{-i (H_g+U_r) n})$. {\it $S_r$ is an adiabatic invariant. It serves as the key parameter, analogous to the Earth's number of atoms in our opening paragraph, controlling the gravitational ensemble.} 


We perform the gravitational trace Tr$_g$ in the semiclassical (small $\hbar$) approximation.  For general relativity coupled to radiation and a cosmological constant, the exponent in the path integral is
\begin{align}
iS/\hbar= i L_{Pl}^{-2} V \int dt \left( -3 {\dot{a}^2\over n} +n(3 \kappa a^2 -\lambda a^4 - r)\right).
 \label{e0}
\end{align}
Here, $L_{Pl} \equiv (8 \pi G \hbar)^{{1\over 2}}$, $ \dot{a}\equiv da/dt$ and $V$ is the comoving volume. The proper radiation density is $\rho_r=L_{Pl}^{-2} r/a^4$ (with $r\equiv U_r L_{Pl}^2/V$) and the dark energy density is $\rho_\Lambda=L_{Pl}^{-2}\lambda$. ($r$ has dimensions of length squared, $\lambda$ of inverse length squared, $a$ of length and $n$ is dimensionless.)  Since the exponent is odd under  $n\rightarrow -n$, saddles occur in pairs with equal and opposite semiclassical exponents. 


\begin{figure}
     \centering
              \includegraphics[width=8.5cm]{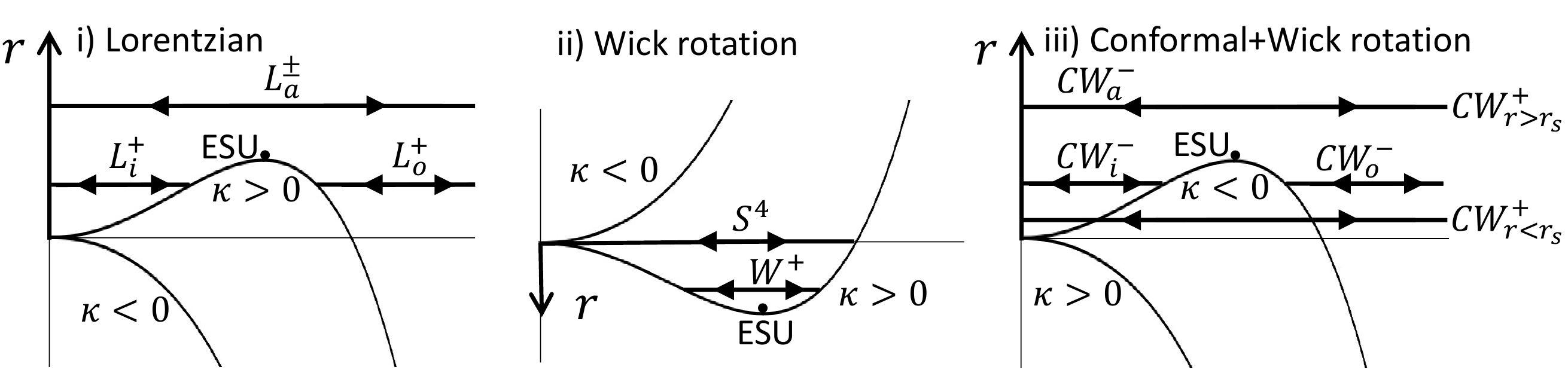}
              \caption{Cosmological solutions for:  i) $a(t)$ in Lorentzian time (see Eqs. (\ref{e0i}),(\ref{e1}), with $n$ real);  ii) $a(t)$ after a Wick rotation $W$, setting $n=-i N$, with $N$ real; iii) $b(t)$ after a conformal {\it and} a Wick rotation $CW$, setting $a(t)=i b(t)$, $n=-i N$ with $b(t)$ and $N$ real.}
 \label{fig:scalefacplot}
\end{figure}

The equations of motion following from (\ref{e0}) are:
\begin{subequations}
  \label{e1}
\begin{align}
\label{e1a}
 3 \left(\dot{a}/n\right)^2+3 \kappa a^2 -\lambda a^4&=r, \\  
 \label{e1b}
 3\ddot{a}/n^2 +3 \kappa a - 2\lambda a^3 &=0,
\end{align}
\end{subequations}
respectively the Friedmann equation and the trace of the Einstein equations, $R=4 \lambda$, to which conformal invariant matter, like radiation, does not contribute. The set of Lorentzian cosmologies, described by $\kappa$, $r$ and $\lambda$ are illustrated in Fig.~\ref{fig:scalefacplot} i). The evolution of the scale factor in  (\ref{e1a}) is that of a particle moving in a potential, with $r$ signifying its energy. For $\kappa>0$ and $r=9 \kappa^2/(4 \lambda)\equiv r_s$, we have Einstein's static universe (ESU) (which \cite{Gibbonsesu} shows is thermodynamically stable). 
For $\kappa<0$, $r>0$ and $\kappa>0$, $r>r_s$, solutions $L_a^\pm$ (for $\kappa \gtrless 0)$ run ``above" the potential.  For $\kappa>0$, $r<r_s$, the ``inner" and ``outer" solutions are $L_i^+$ and $L_o^+$. 

\begin{table*}[t]
\centering
\begin{tabular}{c rrrrrrr}
Case &scale factor $a(t)$ or $b(t)$\qquad $ $& $N\qquad \qquad$ & $S_g/ S_\Lambda\quad \qquad\quad$ &$m \,$  & $z_\pm \quad$ \\ 
\hline 
$W^+$ &$\sqrt{3 \kappa\over \lambda (1+m)} {\rm dn} (\sqrt{\kappa\over 1+m} N t,1-m)$ & $+\sqrt{1+m\over \kappa} 2K(1-m)$ & $+{(1+m)E(1-m)- 2m K(1-m) \over (1+m)^{3/2}}$ &$e^{-\alpha}$ &$e^{\pm \alpha/4}$ \\ 
$CW_{r< r_s}^+$ & $i \sqrt{3 \kappa m \over \lambda (1+m)} {\rm sn} (i \sqrt{\kappa\over 1+m} N t,m)\quad$ & $+ \sqrt{1+m\over \kappa} 2K(1-m)$ & $+{ (1+m)E(1-m) -2 m K(1-m)\over (1+m)^{3/2}}$ & $e^{-\alpha}$  &$e^{\pm \alpha/4}$  \\
$CW_{r> r_s}^+$ & $i \sqrt{3 \kappa m \over \lambda (1+m)} {\rm sn} (i \sqrt{\kappa\over 1+m} N t,m)\quad$ & $- \sqrt{1+m\over \kappa} 2K(1-m)$ & $-{ (1+m)E(1-m) -2 m K(1-m)\over (1+m)^{3/2}}$ & $e^{i\theta}$ &$e^{\pm i\theta /4}$  \\
$CW_i^-$ & $ \sqrt{3 |\kappa| m \over \lambda (1+m)} {\rm sn} (\sqrt{|\kappa|\over 1+m} N t,m)\quad$ & $-\sqrt{1+m\over |\kappa|} 2K(m)\qquad$ &  $2{(1+m)E(m)-(1-m)K(m) \over (1+m)^{3/2}}$ & $e^{-\alpha}$  & $\pm i e^{-\alpha/4}$ \\
$CW_a^-$ & $ \sqrt{3 |\kappa| m \over \lambda (1+m)} {\rm sn} (\sqrt{|\kappa|\over 1+m} N t,m)\quad$ & $-{\rm Re}\left(\sqrt{1+m\over |\kappa|} 2K(m)\right)$ &  ${\rm Re}\left(2{(1+m)E(m)-(1-m)K(m) \over (1+m)^{3/2}}\right)$ & $e^{i \theta}$  &$\pm i e^{\mp i \theta/4} $ \\

\hline
\end{tabular}
 \caption{Analytic forms for Euclidean instantons (Fig.~\ref{fig:scalefacplot}, ii) and iii)), periodic in $t$ with period unity, from which the Lorentzian solutions (Fig.~\ref{fig:scalefacplot}, i)) are obtained by analytic continuation.
Here, sn$(z,m)$, dn$(z,m)$ are Jacobi elliptic functions and $K(m)$, $E(m)$ are complete elliptic integrals~\cite{AbramowitzStegun}. $W$ denotes Wick rotation and $CW$ combined conformal and Wick rotations. Superscript $\pm$ indicates $\kappa \lessgtr 0$; $m$ satisfies $m/(1+m)^2= \lambda r/(3 \kappa)^2$. For  $0<r\leq r_s$, $m=e^{-\alpha}$ with $0\leq\alpha<\infty$, and for $r\geq r_s$, $m=e^{i\theta}$ with $0\leq\theta<\pi$. 
$z_\pm$ are the two branch points pictured in Fig.~\ref{fig:branchcuts}.}
\label{tab:cosinsts}
\end{table*}


The line element (\ref{e0i}) may be rendered Euclidean in two ways: i) $W$, the usual Wick rotation, which sets  $n=-iN$ with $N$ real and ii) a $CW$ rotation combining $W$ with a conformal rotation $a(t)=ib(t)$, with $b(t)$ real. The first yields a metric with ``all positive" signature, the second ``all negative." As we shall now explain, the second rotation is more generally applicable in cosmology.  It is easily seen from  Fig. \ref{fig:scalefacplot}, ii), that $W$ only yields cosmological instantons for positive $\kappa$ and $r\leq r_s$. These cosmologies are intermediate between de Sitter and the ESU, both of which have static descriptions. Hence, they may be considered as ``close to equilibrium." On the other hand, the $CW$ rotation yields instantons for all $\kappa$, positive or negative, and positive $r$ and $\lambda$. Since all these cosmologies, except the ESU, have a horizon and a Hawking temperature~\cite{BT1}, they presumably have a gravitational entropy. The $CW$ rotation yields an instanton with positive entropy, for {\it all} parameter values: where there is overlap with the $W$ instantons, the two entropies agree. 

The $CW$ rotation actually follows from the boundary conditions we proposed in Refs.~\cite{BFT1,BFT2, BT1}. There, we considered scale factors with zeros, {\it i.e.}, conformal singularities, of a special type. All variables are classified according to whether they are even or odd about the conformal zero, and we insist upon analyticity there. In particular, $a(t)$ is odd so it has a simple analytic zero at $t=0$ and becomes imaginary when we Wick rotate about $t=0$. So, as well as continuing $n$ to $-i N$ with $N$ real, we set $a(t)=i b(t)$ with $b(t)$ real. Conformal invariant matter, which is all we consider here, does not couple to $a(t)$ at all and hence it is insensitive to the conformal rotation. However, the gravitational action (\ref{e0}) {\it is} affected by it. For $\kappa>0$ and $r<r_s$, the cases intermediate between de Sitter and ESU, we find a positive semiclassical exponent requires $N>0$, the usual sign. However, for $\kappa<0$ or for $\kappa>0$ and $r>r_s$,  a positive exponent requires $N<0$. The sign of $N$ indicates the orientation of Euclidean time with respect to the spacelike slices. Hence, our Lorentzian ensemble uses two {\it different} Euclidean descriptions: one for the radiation and one for gravity. We believe this to be reasonable since we regard the Lorentzian description as fundamental and the combined radiation/gravity ensemble is far from equilibrium, with the radiation temperature $T_r$ totally distinct from the Hawking temperature. We emphasize that our new instantons would {\it not} be accessible if one insisted upon using a single Wick rotation to describe these coupled systems, at totally different temperatures. Indeed, our work seems to indicate that taking a Euclidean description to be fundamental (see, for example, the recent approach of Ref.~\cite{KS}, reviewed in Ref.~\cite{Witten}) may {\it preclude} one from describing a realistic cosmology.

\begin{figure}
     \centering
              \includegraphics[width=8cm]{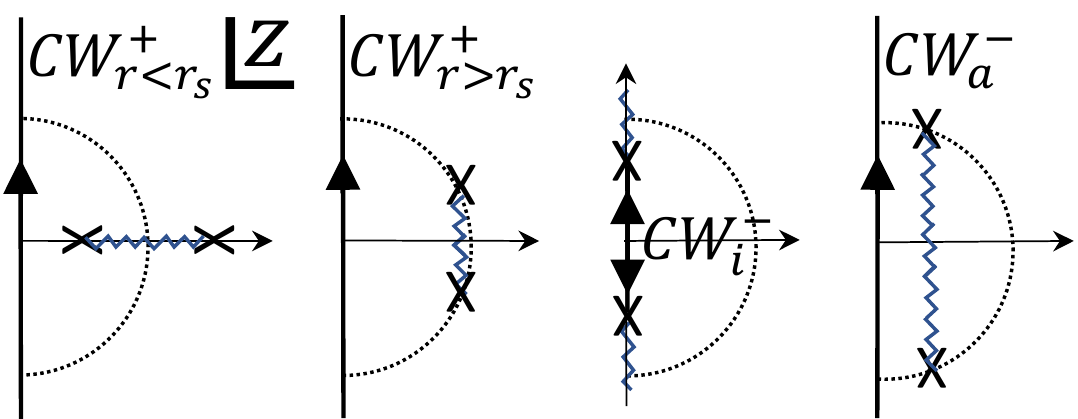}
              \caption{Branch cuts in the complex $z$-plane (see (\ref{e3})). In each case, only two of four branch points are shown (relative to the unit circle, shown dotted).}
 \label{fig:branchcuts}
\end{figure}

Our new instantons are displayed in Table \ref{tab:cosinsts}. For $r>0$, they are periodic in $0<t\leq 1$, with topology $S^3\times S^1$ or (compactified $H^3)\times S^1$.  We choose $N$ so the period in $t$ is unity and the semiclassical exponent is positive. In fact, one can solve (\ref{e1a}) for $\dot{a}/N$ and substitute back into (\ref{e0}). Upon rescaling $a$ via $a=c z$, one obtains
\begin{align}
iS\propto \int dz \sqrt{(z^2-z_-^2)(z_+^2-z^2)}.
 \label{e3}
\end{align}
 where $c$ is chosen so that $z_+ z_-=1$. The integral runs over a closed contour, representing one period of the solution. The lapse $n$ has been eliminated but the orientation of the $z$ integration contour on the integrand's Riemann surface gives the same sign ambiguity. The Laurent series of the integrand about $z=\infty$ contains only even powers. So, if one analytically continues to avoid that singularity, there is no contribution associated with it. Hence, the contour may be deformed to run between two branch points, shown in Fig.~\ref{fig:branchcuts} and in the final column of Table \ref{tab:cosinsts}. The behavior of the solutions as a function of the parameters is most easily understood from this picture.  
 
 For example, the $W^+$ instanton and the $CW_{r<r_s}^+$  instantons are defined by the same branch points (see leftmost panel of Fig.~\ref{fig:branchcuts}). For the $W^+$ instanton, the solution runs around the cut (just above and then just below it) on the real $z$-axis. For the $CW_{r<r_s}^+$  instanton, we deform the contour from the imaginary $z$-axis into the right half $z$-plane and around the cut, obtaining the same result. 
As $r$ is increased to $r_s$, the branch points annihilate and the integral vanishes, as expected since the ESU has no horizon. Increasing $r$ further, branch points appear above and below $z=1$, moving apart on the unit circle as $r$ is increased, reaching $z=e^{\pm i \pi/4}$ at infinite $r$.  For $\kappa$ negative and $r<r_s$ we have the $CW_i^-$ and the $CW_o^+$ solutions shown in Fig.~\ref{fig:scalefacplot} iii). They, too, are defined by the same branch points. As the third panel of Fig.~\ref{fig:branchcuts} shows, the solution for $CW_i^-$ runs between the two ``inner" branch points $\pm ie^{-\alpha/4}$. Whereas the solution for $CW_i^-$ runs between the ``outer" branch points $\pm ie^{\alpha/4}$, via the point at infinity. Again, a contour deformation shows that the two semiclassical exponents are equal. For this reason we do not show $CW_o^-$ in Table  \ref{tab:cosinsts}.

 \begin{figure}
    \centering
    \begin{subfigure}[t]{0.15\textwidth}
        \centering
        \includegraphics[width=\linewidth]{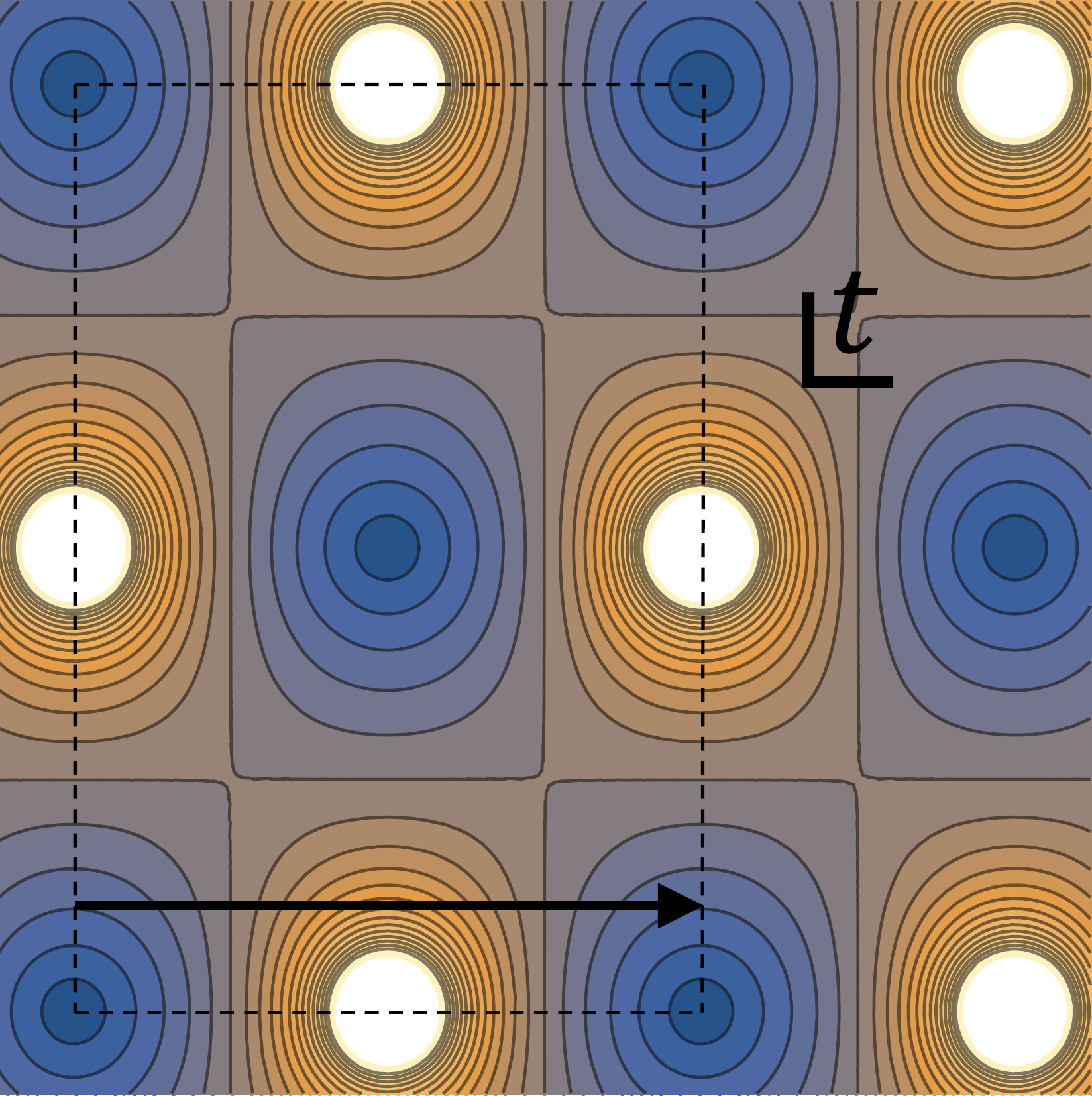} 
    \end{subfigure}
   \hskip .6cm
    \begin{subfigure}[t]{0.18\textwidth}
        \centering
        \includegraphics[width=\linewidth]{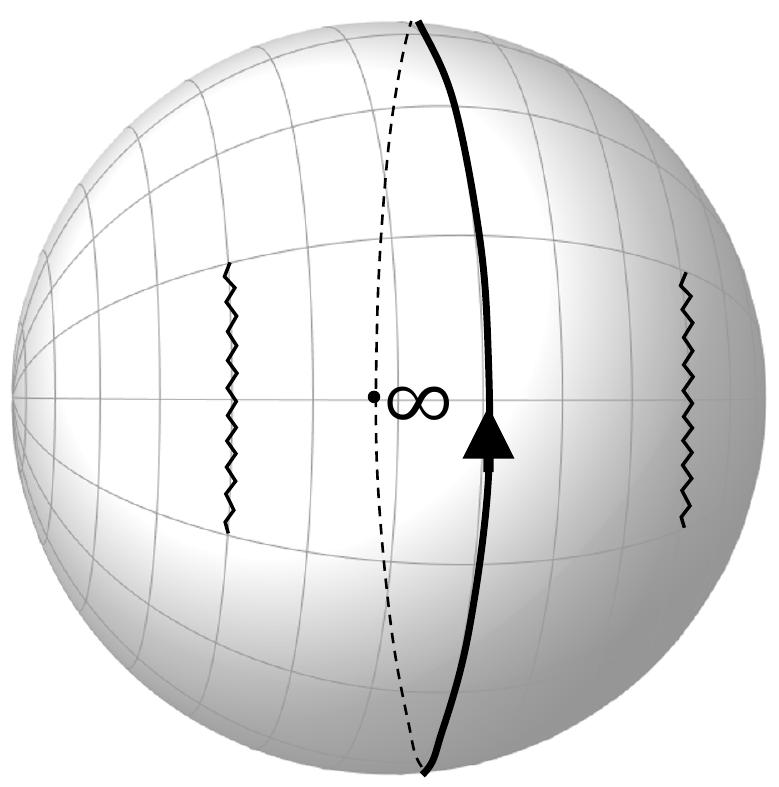} 
    \end{subfigure}
\caption{Double periodicity and topology. Left: contours of $| b(t)|$ in the complex $t$-plane for $CW^+_a$ (Table~\ref{tab:cosinsts}), for $\theta=\pi/2$. Zeros are dark blue, poles white. Right: branch cuts in the integrand of $iS$ on the Riemann $z$-sphere  (see (\ref{e3}) and ensuing text). }
    \label{fig:finite}
\end{figure}

 Fig.~\ref{fig:finite} shows how the complex $t$-plane and $z$-plane pictures are related. The left panel shows the double periodicity of a solution in the $t$-plane. The fundamental domain (shown in dashed lines) is a rectangle with opposite sides identified, {\it i.e.}, a torus. The black vector indicates a period: if the $t$-integral for $iS$ is taken on any contour connecting two points separated by a period, the same value results. Hence $iS$ is a topological invariant.  The right panel shows the Riemann $z$-sphere, with its two square root branch cuts. Since there are two Riemann sheets, a second copy of the Riemann sphere is joined to the first along the right and left edges of the branch cuts. By ``opening" the cuts, one again finds a torus. Since $a(t)$ maps the fundamental $t$-plane domain to the $z$-plane, one to one, the two tori are identified.

\begin{figure}
     \centering
              \includegraphics[width=8cm]{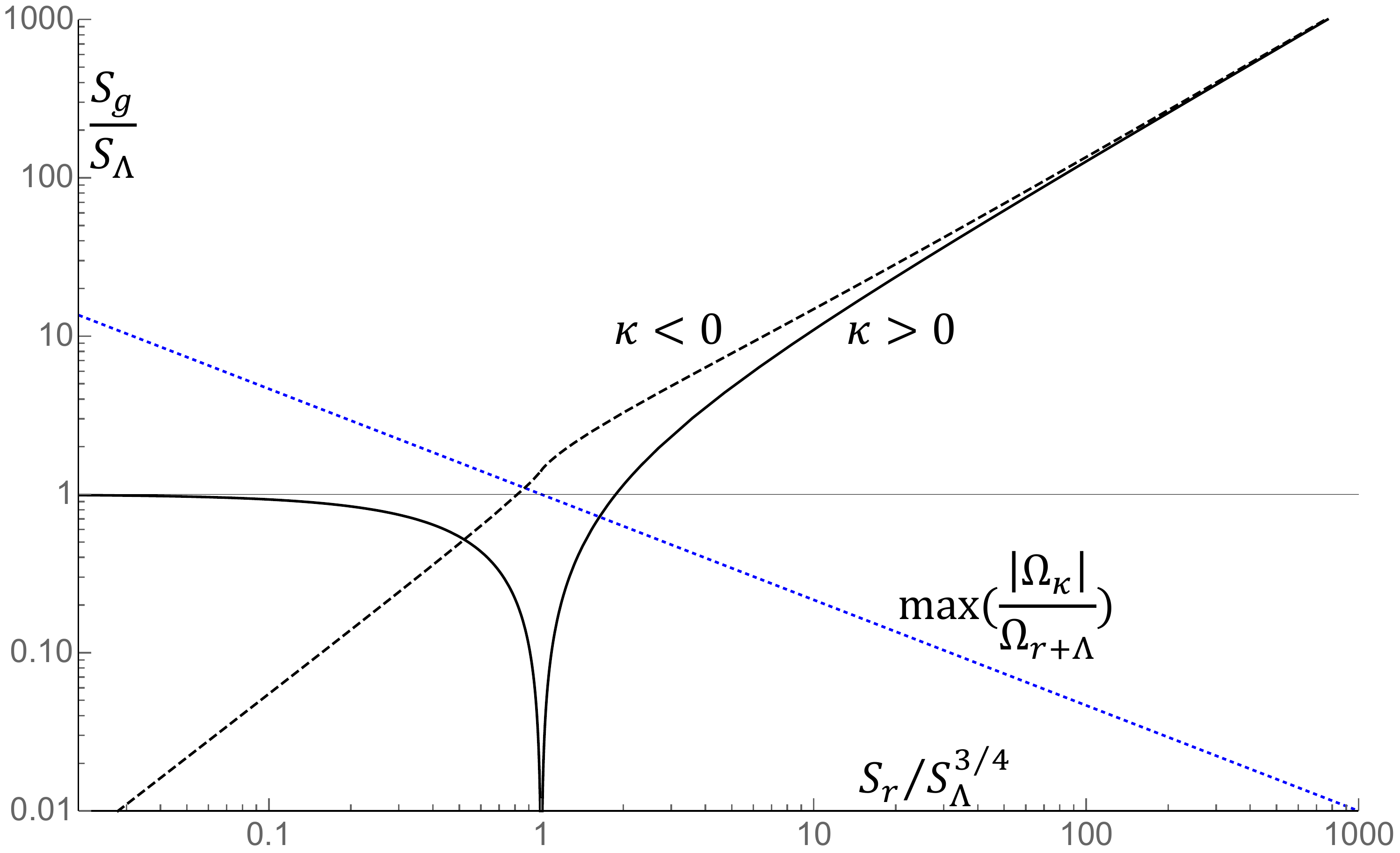}
              \caption{Gravitational entropy $S_g$  as a function of the entropy in radiation $S_r$. The dotted blue line shows the maximum size of the curvature term in the Friedmann equation, relative to the total energy density~\cite{noteent}.}
        \label{fig:entropyplot}
\end{figure}

In the flat limit, $\kappa\rightarrow 0$, the $CW^+_{r>r_s}$ and $CW_a^-$ solutions tend to $b(t)=(r/\lambda)^{1/4} e^{- i\pi/4} {\rm sn} (2 K({1\over 2}) e^{ i\pi/4} t,-1)$. Unsurprisingly, the curvature is dynamically unimportant in the limit. What is slightly less obvious is that the Euclidean solution for $b(t)$ is actually the {\it same} function as the Lorentzian solution for $a(t)$ so the descriptions are self-dual. 
The semiclassical exponent diverges as $V\sim \kappa^{-3/2}$ and we obtain for the gravitational entropy
\begin{align}
S_g=iS/\hbar \approx C V r^{3/ 4} \lambda^{-1/ 4} L_{Pl}^{-2} \approx D S_r S_\Lambda^{1\over 4},
 \label{e4}
\end{align}
where $S_r$ is the total entropy in radiation, $S_\Lambda=24 \pi^2\lambda^{-1} L_{Pl}^{-2}$ is the de Sitter entropy, $C=8 K(1/2)/\sqrt{3}\approx 8.56$ and $D=2^{1/4} \Gamma(1/4)^2/(3^{3/4} \pi) \approx 2.18$~\cite{noteent}. Fig. \ref{fig:entropyplot} shows the gravitational entropy $S_g$, in units of the de Sitter entropy $S_\Lambda$, versus the total entropy in radiation $S_r$ in units of $S_\Lambda^{3\over 4}$, for $\kappa>0$ (solid line) and $\kappa<0$ (dashed line). As mentioned, the gravitational entropy diverges with the volume in the flat limit. Hence, in the regime where $S_g$ dominates the total entropy, the larger the total entropy, the smaller $\kappa$ must be. For example, from Fig. \ref{fig:entropyplot}, if the gravitational entropy exceeds the de Sitter entropy by a thousand, the curvature term in the Friedmann equation is at most a hundredth the size of the combined radiation and dark energy terms.

We now ask whether, in backgrounds where $\Omega_{\kappa}$ is always small, our new instantons represent local maxima of the entropy for observable perturbations.  On wavelengths larger than the Hubble radius, tensor perturbations $h_T$ describe anisotropies. On shorter wavelengths, they describe gravitational waves. To second order, tensors contribute $iS^{(2)}\sim i \int d^4x \sqrt{-\gamma} \, a^2 \left(n^{-1} \dot{h}_T^2- n ({\bf \nabla} h_T)^2\right)$.~Scalar perturbations can be treated as the hydrodynamic modes of a relativistic fluid coupled to gravity (see, {\it e.g.}, Ref.~\cite{MFB}). From their (10.62), setting $v=z \zeta$, we obtain $iS^{(2)} \sim i \int d^4x \sqrt{-\gamma}z^2 \left(n^{-1}\dot{\zeta}^2 -n c_s^2({\bf \nabla} \zeta)^2\right)$. Restoring $n$ in their (10.43b), $z$ is odd under $n\rightarrow -n$ (and, like $a$, odd under $t\rightarrow -t$). Thus, under our $CW$ rotation, both $a$ and $z$ become imaginary.
In our previous papers \cite{BFT1, BT1} we showed that $h_{T}$ and $\zeta$ are both even, so remain real.  
Setting $n=-i N$ with $N<0$, we obtain $iS^{(2)}\leq 0$ in both cases. Finally, consider conformal perturbations  $\delta b(t)$. From (\ref{e0}), $iS^{(2)}\sim - \int_0^1 dt \left({\dot{\delta b}^2\over N} +N(\kappa  +2 \lambda b_0^2 )\delta b^2 \right)$, with $b_0(t)$ the instanton and $\delta b(t)$ periodic in $t$. For $N<0$, a convergent measure requires a negative kinetic term, hence $\delta b$ to be imaginary, as per the arguments of Refs.~\cite{GHP,Hawking}. For $\kappa>0$,  evidently $iS^{(2)}\leq 0$. For  $\kappa<0$ and  $r>r_s$, we find $iS^{(2)}\leq 0$, with a zero mode appearing at $r=r_s$.



Our results point to a new and remarkably economical explanation for the large scale geometry of the universe. However, much remains to be done. We have only sketched the Lorentzian gravity-matter ensemble and its path integral for which we expect our instantons provide relevant saddles. Such an ensemble could be defined by tracing over the initial data. Our $CPT$-symmetric boundary conditions are closely related to those studied in the context of Penrose's ``Weyl curvature hypothesis"~\cite{Penrose, Tod, Anguige}. Newman has shown that classical Einstein gravity coupled to a perfect radiation fluid, with conformal singularities of just the type we consider, has a well-posed Cauchy problem for which the conformal 3-geometry and radiation density on the singular 3-surface provide initial data~\cite{Newman1,Newman2,Claudel}. As in our analysis, all dynamical quantities are either even or odd in the conformal factor, the natural time variable. Of course, the standard model is not exactly conformal in the $UV$. However, we have recently shown that the {\it leading} quantum corrections to the trace anomaly may be cancelled by dimension zero fields, without introducing any new particles. The same fields cancel the leading corrections to the vacuum energy, and could seed scale-invariant large-scale density perturbations~\cite{BT2}.  Finally,  our calculations must be extended to include non-conformal matter, including massive fields and black holes, to provide a more complete analysis of the thermodynamics of the universe.

{\bf Acknowledgements:}  We thank Job Feldbrugge, Stefan Hollands,  Malcolm Perry, Bill Unruh and Roman Zwicky for helpful feedback. The work of NT is supported by the STFC Consolidated Grant `Particle Physics at the Higgs Centre' and by the Higgs Chair of Theoretical Physics. Research at Perimeter Institute is supported by the Government of Canada, through Innovation, Science and Economic Development, Canada and by the Province of Ontario through the Ministry of Research, Innovation and Science.


\begin{thebibliography}{99}
\bibitem{note1}
We only consider cosmologies in which space is compact.
\bibitem{noteent}  For simplicity we approximate the entropy density $s_r$ by $\rho_r^{3/4}$, neglecting modest corrections from numerical factors and the number of degrees of freedom. We also use $K(2)=e^{-i\pi/4} K({1\over2})$, $K({1\over2})= 4 \Gamma({5\over 4})^2/\sqrt{\pi}$.
\end{thebibliography}
\end{document}